\documentclass[reprint,superscriptaddress,amsmath,amssymb,aps]{revtex4-1}
\usepackage[utf8]{inputenc}
\usepackage[T1]{fontenc} 
\usepackage{graphicx}
\usepackage{rotating}
\usepackage{makeidx}
\usepackage{threeparttable}
\usepackage{float}
\usepackage{amsmath}
\usepackage{rotating}
\usepackage{hyperref}

\usepackage{relsize}
\usepackage{etoolbox}
\usepackage[most]{tcolorbox}
\usepackage{hyperref}
\usepackage{dcolumn}
\usepackage{bm}
\usepackage{wrapfig}
\usepackage{float}  
\usepackage[toc,page]{appendix}
\usepackage{gensymb}
\usepackage{epstopdf}
\usepackage{subfiles}
\usepackage{comment} 
\usepackage[autostyle=true]{csquotes} 
\usepackage{makecell}
\usepackage{siunitx}
\usepackage[all]{hypcap} 
\usepackage[most]{tcolorbox}
\usepackage{hyperref}
\usepackage{xcolor}
\usepackage{nicefrac}
\usepackage{chemformula}
\usepackage{ulem}  
\definecolor{gainsboro}{rgb}{0.86, 0.86, 0.86}
\usepackage[backgroundcolor=gainsboro]{todonotes}
\hypersetup{
  colorlinks   = true, 
  urlcolor     = blue, 
  linkcolor    = blue, 
  citecolor   = blue 
}

\newcommand{\weg}[1]{}

\newcommand{\orbs}{ $ d_{x^2-y^2}, d_{3z^2-r^2} $  }

\begin{document}

\title{Optical control of spin-splitting in an altermagnet}

\author{Sangeeta Rajpurohit} 
\email{srajpurohit@lbl.gov}
\affiliation{Molecular Foundry, Lawrence Berkeley National Laboratory, USA}

\author{Revsen Karaalp}
\affiliation{Advanced Light Source, Lawrence Berkeley National Laboratory, USA}

\author{Yuan Ping}
\affiliation{Department of Materials Science and Engineering, University of Wisconsin-Madison, Madison, Wisconsin 53706, United States}
\affiliation{Department of Physics, University of Wisconsin-Madison, Madison, Wisconsin 53706, United States}
\affiliation{Department of Chemistry, University of Wisconsin-Madison, Madison, Wisconsin 53706, United States}

\author{L.Z. Tan}
\affiliation{Molecular Foundry, Lawrence Berkeley National Laboratory, USA}

\author{Tadashi Ogitsu}
\affiliation{Lawrence Livermore National Laboratory, Livermore, USA}

\author{Peter E. Bl{\"o}chl} 
\affiliation{Institute for Theoretical Physics, Clausthal University of Technology, Germany}
\affiliation{Institute for Theoretical Physics, Georg-August-Universität Göttingen, Germany}

\date{\today}

\begin{abstract}
Manipulating and controlling the band structure and the spin-splitting
in the newly discovered class of magnetic materials known as 'altermagnets'
is highly desirable for their application in spintronics. Based on real-time simulations for
an interacting multiband tight-binding model, we propose optical excitations as an
effective way to selectively control the spin-splitting of an altermagnet.
The consistent treatment of electronic interactions and electron-phonon
coupling in the model allows for a systematic study of the effect of
these interactions on the spin-splitting of the altermagnet
in the ground as well as in the excited-state.  Our simulations reveal that optical excitations
modify the band structure and thus lead to significant changes in the spin-splitting within 50 fs. 
The relative spin-splitting in the conduction band grows up to four times in the
optically excited altermagnet. We disentangle the roles of Coulomb $U$ and $J$
in the enhancement of the spin-splitting in the photoexcited state. Our study elucidates the potential
for exploiting optical control of spin-splitting gaps to obtain
desirable properties in altermagnets on the fastest possible timescales. 

\end{abstract}

\maketitle

Altermagnetism, which has emerged as a new class of magnetism
besides ferromagnetism and antiferromagnetism, 
has drawn a lot of attention in condensed matter physics
\cite{Smejkal2022,Smejkal2020,Smejkal2022_2}.  
Altermagnets exhibit properties of both ferromagnets (FMs) and
antiferromagnets (AFMs). Similarly to AFMs, they have zero net
magnetization with local magnetic moments alternating in real space.
However, like FMs, they break Kramers's degeneracy in
reciprocal space in the absence of spin-orbit coupling (SOC).
Instead of spatial translation and spin-inversion, the opposite spin
sublattices in an altermagnet are connected by spatial rotation. 
Due to the breaking of time-reversal symmetry in
momentum space, there are several unconventional anomalous
magnetic responses predicted in altermagnets, such as the anomalous
Hall effect \cite{Smejkal2020,Feng2022,Betancourt2023} and the
magneto-optical Kerr effect \cite{Isaiah2024}, which have
been verified by experiments. 

The spin-splitting  in altermagnets originates from
anisotropic exchange interactions. Many materials with
the potential to host altermagnetism have been identified
\cite{Smejkal2022,Guo2023}. However, the predicted spin-splitting
of most of them is less than 0.50 eV. Few exceptions are
RuO$_2$, CrSb and MnTe. The maximum spin-splitting of 1.4 eV is
predicted for RuO$_2$, followed by CrSb and MnTe with
spin-splittings of 1.2 eV and 1.1 eV, respectively
\cite{Smejkal2022}. 

After the discovery of alternative magnetism, most research efforts
have focused on identifying promising materials. Although spin-splitting
plays a pivotal role in most of the predicted applications
of altermagnet, including spintronics, the possible ways to manipulate
and control the spin-splitting gap in these materials remain largely
unexplored. Advances in ultrafast science have resulted in promising
routes to manipulate and control the properties of materials
\cite{Torre2021,Rajpurohit2022,Rajpurohit2020}. Optical tuning of spin-splitting
gaps can offer exciting opportunities to modify the
altermagnetic properties on ultrafast timescales.

In this study, we demonstrate the realization of altermagnetism
with spin-splitting of d-wave symmetry in the presence of orbital
ordering using a multiband interacting tight-binding (TB) model.
Our study shows that spin-splitting in the altermagnetic ground state strongly depends on
Coulomb $U$, but remains relatively unaffected by Coulomb exchange
$J$. Coulomb $J$ only affects the local magnetic moment. The dependence of
spin-splitting gaps on Coloumb parameters suggests that these can be manipulated
in the altermagnetic state on ultrafast timescales via
intense photoexcitation. Our real-time simulations based on
the TB-model reveal significant changes in the band structure
and spin-splitting of the altermagnet induced following the
optical excitation. Notably, the simulations predict substantial
alterations in spin-splitting in the conduction band,
which develop fully within 75 fs. Furthermore, increasing
the light-pulse intensity further amplifies the spin-splitting.
Our findings demonstrate that optical excitations
offer an effective approach to modify spin-splitting
in altermagnets with correlated electrons.

\begin{figure*}[!thp]
\begin{center}
\includegraphics[width=0.95\linewidth]{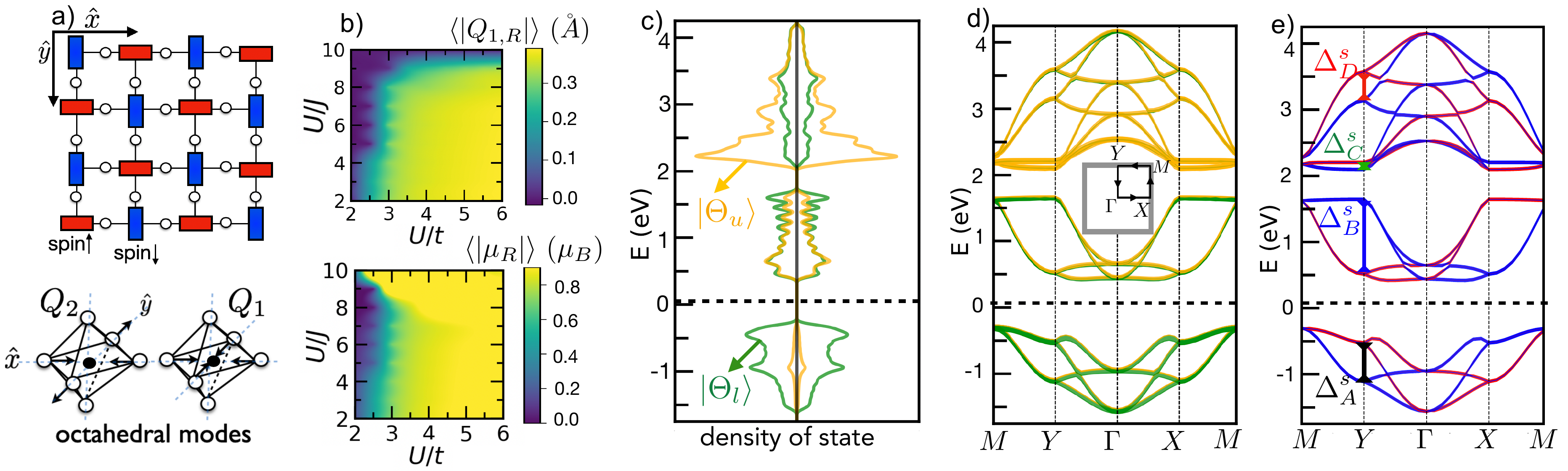}
\end{center}
\vspace{-0.5 cm}
\caption{Altermagnetic ground state of the TB model described in the text
before photoexcitation. (a) Spin and orbital order hosting altermagnetism: 
The rectangles describe the spin and orbital polarization of the TM sites.
Blue and red specify the two opposite spin orientations
directions.  The orientation, horizontal vs. vertical, encodes the orbital polarization
of the $e_g$-electron in the xy-plane along the x and y directions, respectively. 
(b) Average local magnetic moment $\langle|\mu_R|\rangle$ (bottom) and
 octahedral modes $\langle|Q_{2,R}|\rangle$ (top) in the $U/J-U/t$ plane.
(c-d) Density of states (DOS) (c) and band structure (d) projected on local
orbitals $|\Theta_l\rangle$ (green) and $|\Theta_u\rangle$ (orange). 
The left and right DOS shows spin-up and spin-down contributions. 
The thickness of the color represents the intensity of each component in the band
structure. (e) Band structure of the altermagnetic state projected onto spin-up (blue)
 and spin-down (red) components, with the thickness of the colors representing
 the intensity of each component. The band structures are plotted along high symmetry
 points $M(\pi,\pi)-Y(0,\pi)-\Gamma(0, 0)-X(\pi, 0)-M(\pi,\pi)$. The plots in (b) use $U{=}2.50$ eV. 
 The plots c-e use $t=0.833$ eV, $U/t=3.0$ and $U/J=7.0$.}
\label{fig:fig1}
\end{figure*}

\noindent
Firstly, we construct a 2D interacting TB-model that describes the generic features of
altermagnetism driven by orbital ordering in strongly correlated oxides. 
We start with a square planar lattice of corner-sharing oxygen octahedra.
Each such oxygen octahedron hosts a transition-metal (TM) ion in its center.
For each TM site $R$ we consider the d-orbitals of $e_g$ symmetry, $d_{x^2-y^2}$
and $d_{3z^2-r^2}$ and two octahedral distortion modes. 

The potential energy $E_{pot}\Big[\{\psi_{\sigma,\alpha,R,n}\},\{Q_{i,R}\}\Big]$
of the proposed interacting TB-model is expressed in terms of one-electron
wave functions $|\psi_n\rangle{=}\sum\limits_{\sigma,\alpha,R} |\chi_{\sigma,\alpha,R}\rangle \psi_{\sigma,\alpha,R,n}$
and octahedral distortion modes $Q_{i,R}$. The basis set $|\chi_{\sigma,\alpha,R}\rangle$ for
consists of local spin orbitals with spin $\sigma\in\{\uparrow,\downarrow\}$ and orbital
character $\alpha\in$\{\orbs\}. 

The potential energy 
\begin{eqnarray}
E_{pot}&=&E_{e} +E_{ph} +E_{e-ph}
\label{eq:tbm_1}
\end{eqnarray} 
consists of the energy $E_{e}$ of the electronic subsystem, the energy $E_{ph}$ the
phonon subsystems and the electron-phonon (el-ph) interaction energy $E_{e-ph}$. 

\noindent
\textit{Electronic subsystem:} The electronic energy $E_{e}{=}E_{hop}+E_{coul}$ consists
of the kinetic energy $E_{hop}$ and the Coulomb interaction $E_{coul}$. 
The kinetic energy is expressed as 
\begin{eqnarray}
E_{hop}&=&\sum\limits_{R,R',\sigma,n}
f_n \sum\limits_{\alpha,\alpha'} \psi_{\sigma,\alpha,R,n}T_{\alpha,\alpha',R,R'}\psi^*_{\sigma,\alpha',R,n}
\label{eq:tbm_2}
\end{eqnarray} 
where the hopping-matrix elements $T_{\alpha,\alpha',R,R'}$ contribute only onsite and
nearest-neighbor terms between the TM-sites. The $f_n$ are the occupations of the
one-particle wave functions $|\psi_n\rangle$. The hopping matrix elements along
$x$ and $y$ directions are defined as  
\begin{eqnarray}
T^{x}_{R,R'}&=&
-\frac{1}{4}t_{hop}
\left(\begin{array}{cc}
  3   & -\sqrt{3} \\
  -\sqrt{3}   & 1
\end{array}
\right)
\\
T^{y}_{R,R'}&=&
-\frac{1}{4}t_{hop}
\left(\begin{array}{cc}
  3   & +\sqrt{3} \\
  +\sqrt{3}   & 1
\end{array}
\right)
\end{eqnarray}

The Coulomb interaction $E_{coul}$ 
\begin{eqnarray}
E_{coul}&=&
\frac{1}{2}(U-3J_{xc})\sum_{R}
\left(\sum_{\sigma,\alpha}
\rho_{\sigma,\alpha,\sigma,\alpha,R}\right)^2 \nonumber \\ &-&\frac{1}{2}(U-3J_{xc})\sum_R
\sum_{\sigma,\alpha,\sigma',\beta}
|\rho_{\sigma,\alpha,\sigma',\beta,R}|^2 \nonumber \\
&+&\frac{1}{2}J_{xc}\sum_R\sum_{\sigma,\sigma'}(-1)^{\sigma-\sigma'}\sum_{k\in\{x,z\}}
\nonumber\\
&\times&\Bigl[
\Bigl(\sum_{\alpha,\beta}\rho_{\sigma,\alpha,\sigma',\beta,R}
\sigma^{(k)}_{\beta\alpha}\Bigr)
\Bigl(\sum_{\alpha,\beta}\rho_{-\sigma,\alpha,-\sigma',\beta,R}
\sigma^{(k)}_{\beta\alpha}\Bigr)
\nonumber\\
&&
+\Bigl(\sum_{\alpha}\rho_{\sigma,\alpha,\sigma',\alpha,R}\Bigr)
\Bigl(\sum_{\alpha}\rho_{-\sigma,\alpha,-\sigma',\alpha,R}\Bigr)\Bigr],
\end{eqnarray}
is conveniently expressed in terms of the on-site terms of the one-particle-reduced density matrix
${\hat{\rho}}$ with the matrix elements defined as 
\begin{eqnarray}
\rho_{\sigma,\alpha,R,\sigma',\alpha',R}{=}\sum\limits_{n} f_n\psi_{\sigma,\alpha,R,n}\psi^*_{\sigma',\alpha',R,n}.
\end{eqnarray}
 \textit{Phononic subsystem:}
The phononic subsystem consists of two octahedral modes of oxygen around
every TM-site $R$: a Jahn-Teller (JT) mode $Q_{2, R}$ and
a breathing mode $Q_{1,R}$ \cite{Sotoudeh2017}. These phonon modes are defined
by the displacement vectors of oxygen ions from their equilibrium positions.
The JT mode is an asymmetric expansion
of an octahedron in the plane, i.e. expansion along $x$ and contraction along the $y$-direction.
The breathing mode is the isotropic expansion of the oxygen octahedron.  

The restoring energy of this phononic
sub-system $E_{e-ph}$ is expressed as 
\begin{eqnarray}
E_{ph} &=& \sum_{R}
\Big( \frac{1}{2}k_{JT}Q^2_{2,R}
+ \frac{1}{2}k_{br}Q^2_{1,R}\Big)
\end{eqnarray}
where $K_{JT}$ and $K_{br}$ are the restoring force constants of
the JT and breathing mode. Due to the shared oxygen ions, the octahedral
distortions are highly cooperative.

\noindent
\textit{Electron-phonon coupling:} 
We consider strong coupling between the eg-electron at TM-sites to the local
modes $Q_{1,R}$ and $Q_{2,R}$ \cite{Sotoudeh2017} (see SI). 
In 3d TM oxides like manganites and nickelates, this type of el-ph interaction is
another mechanism besides the Coulomb interaction, which causes long-range orbital
ordering \cite{Yunoki2000,Sen2010,Rajpurohit2020}. The el-ph coupling $E_{el-ph}$ in the
model is defined as
\begin{equation}
E_{e-ph} =
\sum_{R,\sigma}\sum_{\alpha,\beta} \rho_{\sigma,\alpha,\sigma,\beta,R}
M^Q_{\beta,\alpha}(Q_{1,R},Q_{2,R}).
\end{equation}
Here $g_{JT}$ and $g_{br}$ are the el-ph coupling constants and $\mathbf{M}^Q(Q_{1,R},Q_{2,R})$ is defined as 
\begin{eqnarray}
\mathbf{M}^Q(Q_{1,R},Q_{2,R})=
\left(\begin{array}{cc}
-g_{br}Q_{1,R} & 
g_{JT}Q_{2,R}\\
g_{JT}Q_{2,R} & 
-g_{br}Q_{1,R}
\end{array}\right)
\end{eqnarray}

The TB-model can describe the strong inequivalent local crystal environment
caused by local el-ph coupling and electronic interactions for two sublattices with
opposite spins, which reduces the symmetry as needed for alternating spin-splitting
of the energy bands in reciprocal space.  

We take model parameters from our previous study
as a reference \cite{Sotoudeh2017,Rajpurohit2020_2, Rajpurohit2020}.
The reference model parameters are  $g_{br} = 2.988$ eV/Å, $K_{br} = 10.346$
eV/Å$^2$, $g_{JT} = 2.113$ eV/Å, $K_{JT} = 5.173$
eV/Å$^2$, $U/t=4.29$, $U/J=2.63$ and $t_{hop} = t = 0.585$ eV. In the present study,
we are interested in the effect of electronic interactions and correlations on the
altermagnetic properties, we vary $U$, $J$ and $t_{hop}$ while keeping the other
parameters fixed.

\noindent
\textit{Altermagnetic ground-state:}
We investigate the region of phase diagram 
with the altermagnetic ground-state, shown in
Figure \ref{fig:fig1}-a. We consider one electron per TM-site.
Figures \ref{fig:fig1}-b displays 
the changes in the average local mode $\langle Q_{1,R}\rangle$ 
and the magnetic moment $\langle|m_R|\rangle$ at the TM-sites
in the altermagnetic ground-state as a function of $U/t$ and
$U/J$ for $U{=}2.50$ eV. For $U/J{<}2.5 $ and $U/t{<}2.0$, 
the system exhibits a non-magnetic metallic phase characterized by
TM-sites with zero magnetic and orbital moment. As $U/J$ and $U/t$
increase, the system undergoes
metallic to an insulating altermagnetic phase transition. In the
altermagnetic phase, the TM-sites 
have finite magnetic moments and eg-orbital polarization. 
These local magnetic moments are arranged in
antiferromagnetic (AFM) order. The local eg-orbital polarization 
forms long range staggered orbital ordering as shown in Figure
\ref{fig:fig1}-a. This altermagnetic state remains stable for
$2.5{<}U/J{<}6.0$ and $2.0{<}U/t{<}6.0$.

Figures \ref{fig:fig1}-c show the
density of states of the altermagnetic
ground-state. In the ground state, all TM-sites
have one eg-electron and are JT active. The JT-effect
lifts the degeneracy of the local $\rm{e_g}$-orbitals. 
The lower filled $\rm{e_g}$-state  $|\Theta_{l,R}\rangle$ located
between -1.5 and 0 eV is described by the linear combination
\begin{eqnarray}
|\Theta_{l,R}\rangle = -\sin{(\gamma)}|d_{x^2-y^2}\rangle \pm \cos{(\gamma)}|d_{3z^2-r^2}\rangle
\label{eq:eg_1}
\end{eqnarray} 
of the $\rm{e_g}$-orbitals $d_{x^2-y^2}$ and $d_{3z^2-r^2}$ at site
$R$ with $\gamma{=}60^{\circ}$. 
The corresponding unoccupied state $|\Theta_u\rangle_R$ are 
\begin{eqnarray}
|\Theta_{u,_R}\rangle = \cos{(\gamma)}|d_{x^2-y^2}\rangle \pm \sin{(\gamma)}|d_{3z^2-r^2}\rangle.
\label{eq:eg_2}
\end{eqnarray} 
The upper sign in front of the $|d_{3z^2-r^2}\rangle$-state
describes an orbital polarization along $x$ and the lower sign
describes an orbital polarization along the $y$ direction.

Figure \ref{fig:fig1}-e shows the spin-polarized band structure of
the calculated altermagnet state. Clearly, the spin degeneracy of the
bands is lifted in momentum space, with the largest spin-splitting
gap at high-symmetry points $X{=}(\pi, 0)$ and $Y{=}(0, \pi)$. 
The spin-splitting is present for both the valence band and the conduction band. 
The magnitude of the spin-splitting for the valence band,
indicated by $\Delta^s_A$ in Figure \ref{fig:fig1}-e (black) is
approximately 0.57 eV. The spin-splitting of $1.10$ eV
in the lower conduction band, indicated by $\Delta^s_B$ (blue), is significantly higher
than the corresponding values of $\Delta^s_C=0.07$ eV (green) and $\Delta^s_D=0.45$ eV (red)
in the upper conduction band. The spin-polarized bands
\textcolor{black}{with opposite spins} are related by $\pi/2$
rotations in the momentum space. 

\begin{figure}[!thp]
\begin{center}
\includegraphics[width=\linewidth]{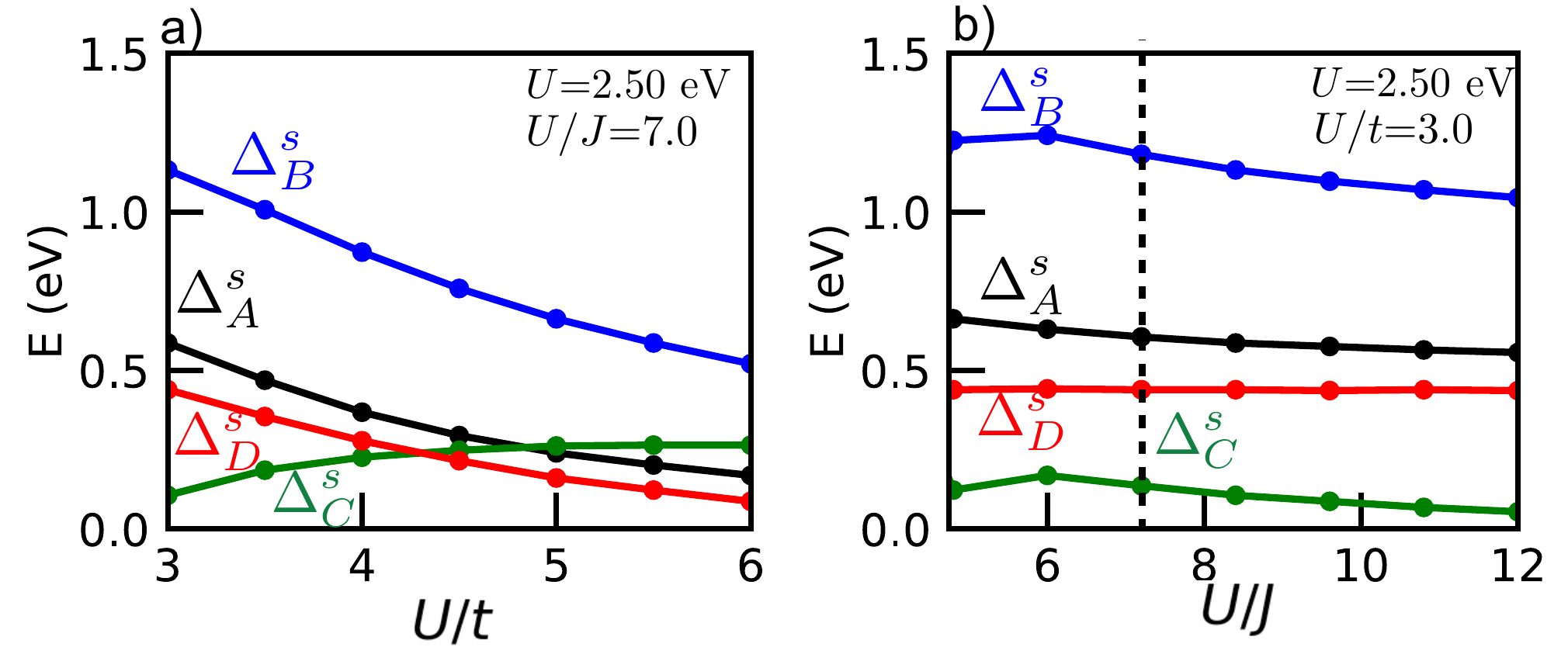}
\end{center}
\vspace{-0.5 cm}
\caption{Spin-splitting gaps in the altermagnetic ground state as functions
of $U/t$ (a) and $U/J$ (b). The color codes are the same as \ref{fig:fig1}-e. 
The other model parameters are set to reference values, which are
specified in the main text. The vertical dashed line in (b) indicates $U/J{=}7.0$ 
used to study photoexcitation.}
\label{fig:fig2}
\end{figure}

Figure \ref{fig:fig2}
shows the magnitude of the spin-splitting gap as a function of $U/t$ and $U/J$. 
The spin-splitting gap $\Delta^s_A$ of the valence band and $\Delta^s_B$ of
the lower conduction band decreases with $U/t$. The spin-splitting in
the upper conduction band $\Delta^s_D$
increases with $U/t$, while $\Delta^s_C$ decreases with $U/t$. 
The spin-splitting gaps remain largely unaffected, while the
magnetic moments are weakly affected by $U/J$.

The above study of the phase diagram suggests that the properties of the
altermagnetic state in correlated electrons systems are sensitive
to the Coulomb $U$ and $J$ parameters. The strength of electronic
interactions and correlations can be altered through material composition
or external means, such as applying an electromagnetic field or varying
the temperature. In the following sections, we use real-time simulations
based on the proposed TB-model to demonstrate optical excitations as
an effective way to modify the properties of altermagnet on ultrafast
timescales.

\noindent
\textit{Photo-excitation study with rt-TDDFT:}
To study the evolution of spin-splitting under photo-excitation 
in the altermagnetic state described above, we employ a simulation
framework similar to real-time time-dependent density-functional
theory (rt-TDDFT) formalism \cite{Runge1984} based on the model defined in Eq.\ref{eq:tbm_1}.
The one-particle wavefunctions of eg-electrons are propagated according to
the time-dependent Schrödinger equation. For the sake of simplicity, the atoms
are kept frozen while studying the photoexcitation. The photoexcitation is
described by an explicit time-dependent vector potentials $\vec{A}(t)=\vec{e}_s \omega \rm{Im}(A_oe^{-i\omega t})g(t)$
which is coupled to the electrons using Peierls substitution method \cite{Peierls1933}.
Here, $A_o$ is the amplitude of the vector potential and 
$\omega$ is angular frequency.
The envelope $g(t)$ of the light pulse is a Gaussian with a width of 30 fs. 
The polarization $\vec{e}_s$ of light is along the $\hat{x}+\hat{y}$ or the
$\hat{x}-\hat{y}$ direction. The unit cell consists of
$N{=}4$ TM-sites and 8 oxygen atoms. The simulations are performed
with periodic boundary conditions on a 24x24 k-point grid.

\begin{figure}[!htp]
\begin{center}
\includegraphics[width=\linewidth]{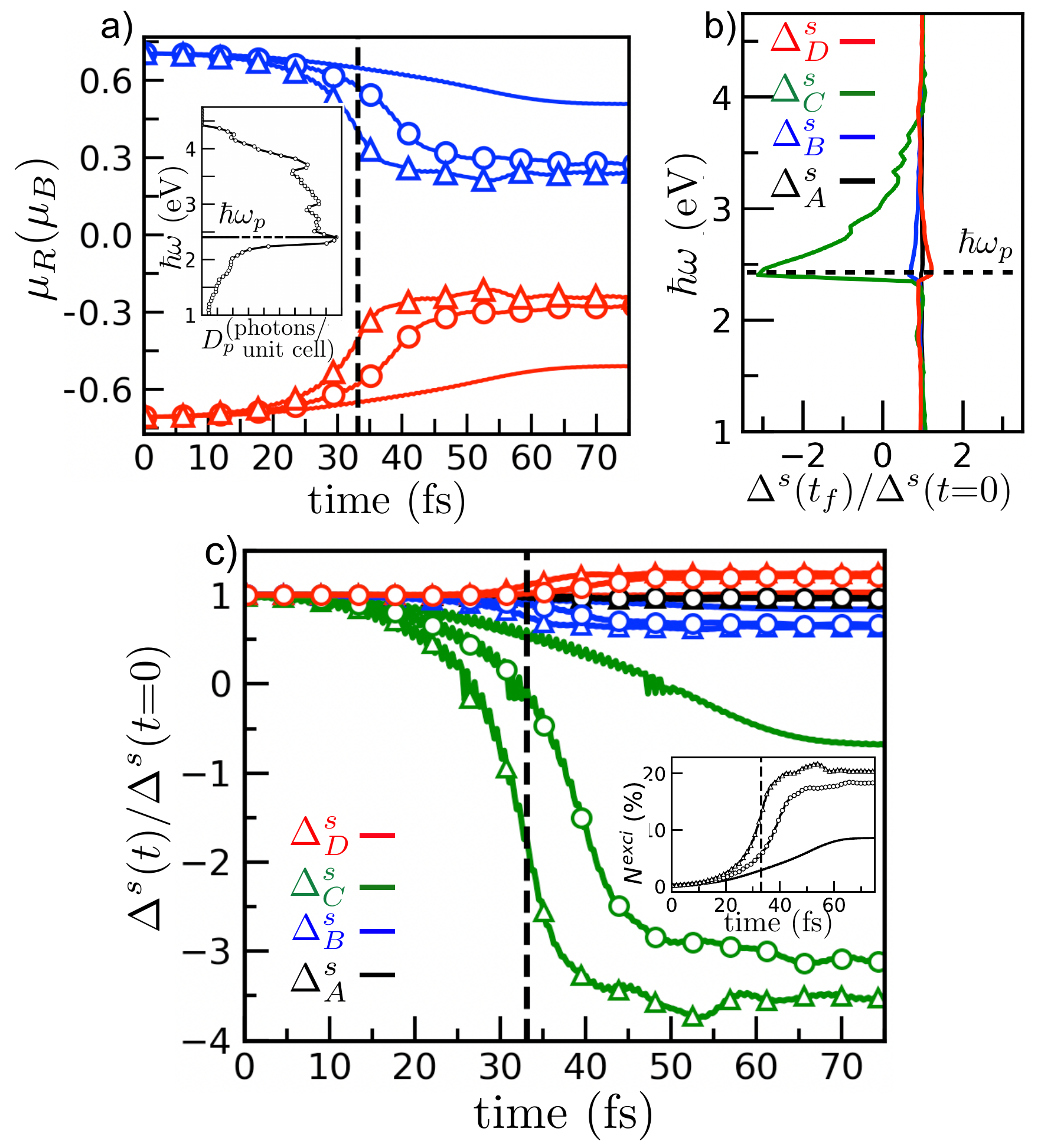}
\end{center}
\vspace{-0.5 cm}
\caption{ (a) Evolution of the local magnetic moment of the
spin-up (blue) and spin-down (red) TM-sites during and after
light-pulse with photon energy $h\omega_p=2.39$ eV 
for different $N^{exci}$. The inset shows the
photon absorption density $D_p$ versus the excitation
energy $\hbar\omega$, where the horizontal line indicates $h\omega_p=2.39$. 
(b) Relative change in spin-splitting $\Delta^s(t_o)/\Delta^s(t{=}0)$ at
the high-symmetry point X at $t_f{=}50$ fs. (c) Evolution of percentage of
excited electrons (inset) and $\Delta^s(t)/\Delta^s(t{=}0)$ (main) at
X-point after photoexcitation for various light intensities. Lines with
triangles, circles, and without symbols correspond to
$N^{exci}{=}$ 8.50\%, 18.30\% and 20.30\%, respectively, after $t_f$=50 fs
in both figures b and c. The vertical line marks the zero point of the
30-fs Gaussian light pulse. The model parameters are consistent with
the panel (e) in Figure 1.}
\label{fig:fig3}
\end{figure}

\noindent
\textit{Absorption Spectra:}
The optical absorption of the altermagnetic state as a function of photon
energy $\hbar\omega$ is shown in the inset of Figure~\ref{fig:fig3}-a. 
It is expressed in terms of the photon-absorption density $D_p$, i.e. the total
number of photons absorbed per site. The photon-absorption density is computed
from energy difference before and after the light pulse as
$D_p{=}\delta E_{pot}/{\hbar\omega N}$ with $E_{pot}$ from Eq.~\ref{eq:tbm_1}.
We attribute the broad absorption peak around $\hbar\omega_{p}{=}2.39$ eV to
dipole-allowed inter-site electronic transitions from majority to minority spin
orbitals located on nearest-neighbor TM-sites belonging to opposite spin sublattices.

\noindent
\textit{Spin-splitting in the photo-excited state:} Let us examine the changes of
the spin-splitting gaps upon optical excitation of the altermagnet. 
We select the photon energy $\hbar\omega_p{=}2.39$ eV with the maximum optical absorption.

For the discussion ahead, we discriminate between two sets of time-dependent wave functions. 
One set of one-particle wave functions with symbol $|\psi_n(\vec{k},t)\rangle$ defines the
time-dependent Slater determinant. These wave functions are obtained via the \textit{time-dependent}
Schr\"odinger equation with the Hamiltonian $\hat{h}(\vec{k},t)$. The second set of one-particle
wave functions with symbol $|\phi_n(\vec{k},t)\rangle$ is obtained with the same instantaneous
Hamiltonian $\hat{h}(\vec{k},t)$, but using the  \textit{time-independent} Sch\"odinger equation.
The eigenvalues of this Hamiltonian are used to calculate the band structure and spin-splitting gaps $\Delta^s$. 
The instantaneous occupations $f_m(\vec{k},t)$ of the eigenstates of the Hamiltonian are obtained
by projecting the time-dependent wave functions $|\psi_n(\vec{k},t)\rangle$ onto the
eigenstates $|\phi_m(\vec{k},t)\rangle$ of the Hamiltonian, i.e. $f_m(\vec{k},t)=\sum_{q=1}^{N_e}|\langle\psi_q(\vec{k},t)|\phi_m(\vec{k},t)\rangle|^2$.

The ratio of photo-excited electrons  is 
\begin{eqnarray}
N^{exci}=\frac{1}{2}\frac{\sum_{\vec{k}}|f_m(\vec{k},t_f)-f_m(\vec{k},t_i)|}{\sum_{\vec{k}}f_m(\vec{k},t_i)}
\end{eqnarray}
where $t_i$ is a time before and and $t_f$ is a time after the light pulse. 
The spin-splitting gaps $\Delta^s(t)$ are extracted from the eigenvalues
$\epsilon_n(\vec{k},t)$ of the instantaneous electronic Hamiltonian. 
The z-component of the local magnetic moment for site $R$ is 
\begin{eqnarray}
\mu_{z,R}(t){=}\sum_{\alpha,n}\Big(
\Big|\psi_{\uparrow,\alpha,R,n}(t)\Big|^2 -\Big|\psi_{\downarrow,\alpha,R,n}(t)\Big|^2\Big)\;. 
\end{eqnarray}

Figures \ref{fig:fig3}-a and \ref{fig:fig3}-c show the evolution of local magnetic moments $\mu_R$ and spin-splittings
relative spin-splitting $\Delta^s(t)/\Delta^s(t{=}0)$ for light pulses with three different intensities,
corresponding to $N^{exci}{=}$8.50\%, 18.30\%, and 20.30\% photoexcited electrons. 
The optical excitations are charge-transfer transitions between neighboring TM-sites. 
Because the two sites participating in the charge transfer are antiferromagnetic,
the excitation reduces their magnetic moments, which is evident in Figure \ref{fig:fig3}-a (blue and red). 

The spin-splitting gaps respond instantaneously, i.e. during the photoexcitation process:
Two of the spin-splitting gaps, namely $\Delta^s_B$ and $\Delta^s_C$ respond strongly
to optical excitation. For $N^{exci}=18.30\%$ we find a change of 0.35~eV of $\Delta^s_B$
and 0.41~eV  of $\Delta^s_C$. The other spin-splittings are small in comparison with a
change of 0.01~eV in $\Delta^s_A$ and 0.08~eV  in $\Delta^s_D$.

The spin-splitting gap with the largest response to photoexcitation, namely $\Delta^s_C$,
is between states that were nearly degenerate initially. This spin-splitting gap, $\Delta^s_C$,
is in the upper conduction band, where most of the excited electrons are located.
Initially, the spin-splitting $\Delta^s_C$ falls off to zero, and then rises again with
an opposite sign. For the more intense excitations, i.e. with
$N^{\text{exci}} = 18.30\%$ and $20.30\%$, $\Delta^s_C(t)$ becomes nearly
four times its equilibrium value within 50~fs, see Figure \ref{fig:fig3}-c (green).

Figure \ref{fig:fig3}-b shows the relative change in spin-splitting 
$\Delta^s(t_f)$$/\Delta^s(t{=}0)$ at the high-symmetry point X at $t_f{=}50$ fs
as a function of photon energy $\hbar\omega$. The largest changes
$\Delta^s(t_f)/\Delta^s(t{=}0)$ of the spin-splitting gaps occur for photon
energies near the bottom of the broad absorption peak, namely at $h\omega_p$=2.39 eV. 

\begin{figure}[!thp]
\begin{center}
\includegraphics[width=\linewidth]{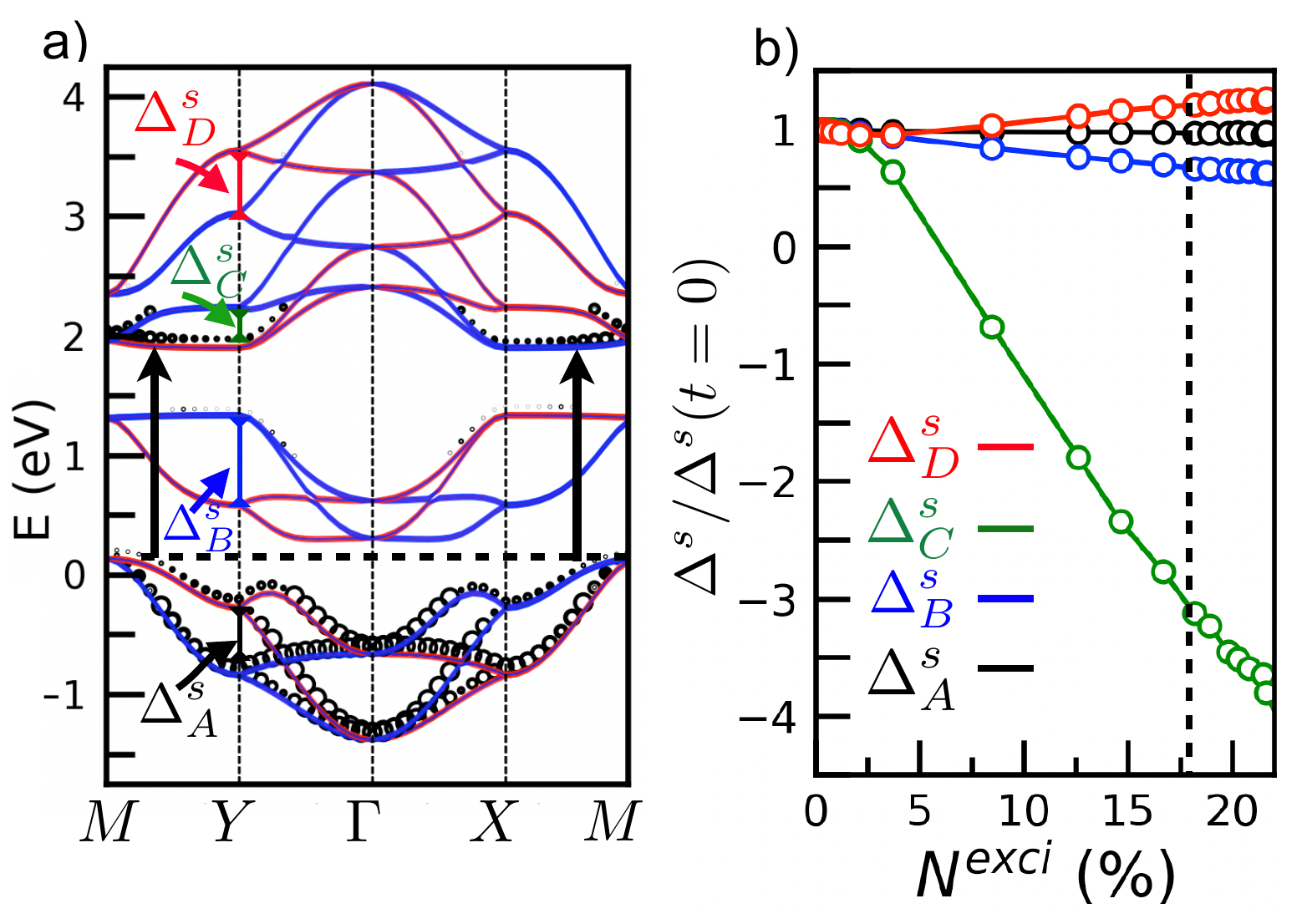}
\end{center}
\vspace{-0.5 cm}
\caption{(a) Band structure of the photoexcited altermagnet at $\hbar\omega_p=2.39$ eV and $N^{exci}{=}18.30\%$, 
showing also the electron distribution in the valence and conduction bands. The size of the
circles represents the occupancy of the electronic levels. The horizontal
dashed line indicates the Fermi level. The vertical arrows indicate the
electronic transitions. (b) Spin-splitting gaps as a function of the
percentage $N^{exci}$ (\%) of photoexcited electrons with $\hbar\omega_p=2.39$ eV. The vertical line indicates $N^{exci}{=}18.30\%$. }
\label{fig:fig4}
\end{figure}

Let us rationalize the findings described above on the level of local orbitals.
For the sake of the argument, we consider here only the diagonal elements of the
one-particle reduced density matrix. The photo excitation with photon energy at
the absorption maximum transfers electrons from the lower, majority spin orbital of
one site to the upper, minority spin orbital of a neighboring site. 
This implies that, on each site, the occupation change $\delta n_{l,\downarrow,R}$ of the
filled orbital equals $-N^{exci}$, while the occupation $\delta n_{u,\uparrow,R}$ of
the upper, minority-spin orbital changes by $+N^{exci}$.
Figure \ref{fig:fig4}-a shows the electron distribution in the photo-excited state
at time $t{=}75$ fs for $N^{exci}=20.30\%$.
The excited electrons predominantly occupy the upper conduction band on the
$X-M$ and the $Y-M$ high-symmetry lines in reciprocal space.

The orbital energies are obtained using Janak's theorem\cite{Janak1978}, as
the derivative of total energy with respect to occupations $n_{\sigma\alpha}$ (see SI for more information). 
Hence, the occupation changes $\delta n_{\alpha,\sigma,R}$ result in a
response $\delta\bar{\epsilon}_{\beta,\sigma',R}$ of the orbital energies of the form
 $\delta\bar{\epsilon}_{\alpha,\sigma,R}=\sum_{\beta,\sigma'}\frac{\partial^2 E_{coul}}{\partial n_{\alpha,\sigma,R}\partial n_{\beta,\sigma',R}} \delta{n}_{\beta,\sigma',R}$.

It turns out that the same orbitals, which experience the largest occupation
changes, experience also the largest energy level shifts, namely $\delta\bar{\epsilon}_{l,\downarrow,R}=-\delta\bar{\epsilon}_{u,\uparrow,R}=(U-2J)N^{exci}$.
The band structure changes according to the k-dependent orbital weights, i.e. $\delta\epsilon_n(\vec{k})=\sum_{\alpha,\sigma,R}|\langle\psi_n(\vec{k})|\Theta_{\alpha,\sigma,R}\rangle|^2 \delta\bar{\epsilon}_{\alpha,\sigma,R}$.

Let us investigate the changes in the spin-splitting gaps as a function of the
photoexcited electron population $N^{exci}$. The most significant changes in
relative spin-splitting with increasing $N^{exci}$ are observed for $\Delta^s_C$,
see Figure \ref{fig:fig4}-b. Remarkably, $\Delta^s_C$ increases up to four times
its initial value at high $N^{exci}$.

Our study demonstrates that spin-splittings of altermagnets can be modified
and controlled by optical excitations. These changes occur on ultrafast time scales.
Promising candidates to realize this effect are 3d transition-metal compounds.
The temporal change in spins-splittings predicted here can experimentally be
probed by time-resolved spin angle-resolved photoemission spectroscopy (SARPES)
\cite{Boschini2024,Sobota2021,Rohwer2011}. 

In conclusion, we demonstrate that optical excitations significantly modify
the spin-splitting gaps in an altermagnet with correlated electrons. 
We study this effect by employing real-time simulations based on an interacting
tight-binding model. The modification in spin-splitting is driven by the
photoinduced changes in the band structure. The magnitude of changes
in spin-splittings is strongly sensitive to the intensity of the light-field. 
Our findings suggest that light fields can be an effective tool for modifying
the altermagnetic properties of strongly correlated systems on ultrafast timescales. 

\textit{Acknowledgments:}
Theory and simulation were supported by the Computational Materials Sciences
(CMS) Program funded by the US Department of Energy, Office of Science,
Basic Energy Sciences, Materials Sciences and Engineering Division. Data analysis
was provided by the User Program of the Molecular Foundry, supported by the Office
of Science, Office of Basic Energy Sciences, of the U.S.
Department of Energy under Contract No. DE-AC02-05CH11231. This work was performed
under the auspices of the U.S. Department of Energy by Lawrence Livermore National
Laboratory under Contract DE-AC52-07NA27344
This work was performed under the auspices of the U.S. Department of
Energy by Lawrence Livermore National Laboratory under Contract DE-AC52-07NA27344.
S.R and T.O. are supported by the Computational Materials Sciences Program
funded by the US Department of Energy, Office of Science, Basic
Energy Sciences, Materials Sciences and Engineering Division.
S.R. was partially supported by funds from the UC National Laboratory Fees Research Program 
of the University of California, Grant Number L25CR9003. 
This work is funded in parts by the Deutsche Forschungsgemeinschaft
(DFG, German Research Foundation) 217133147/SFB1073, projects B03 and C03.

\bibliography{ref}

\end{document}